\documentclass[reprint,prl,showpacs,superscriptaddress]{revtex4-2}
\usepackage{amssymb}
\usepackage{amsmath}    
\usepackage{graphicx}   
\usepackage{verbatim}

\usepackage{color}      
\usepackage{booktabs} 

\usepackage[normalem]{ulem}

\usepackage{color}

\begin{document}
\title{On the new and old physics in the interaction of a radiating electron with the extreme electromagnetic field}

\author{M. Jirka}
\affiliation{ELI Beamlines Centre, Institute of Physics, Czech Academy of 
Sciences, Za Radnici 835, 25241 Dolni Brezany, Czech Republic}
\affiliation{Faculty of Nuclear Sciences and Physical Engineering, Czech
Technical University in Prague, Brehova 7, 115 19 Prague, Czech Republic}

\author{P. Sasorov}
\affiliation{ELI Beamlines Centre, Institute of Physics, Czech Academy of 
Sciences, Za Radnici 835, 25241 Dolni Brezany, Czech Republic}

\author{S. V. Bulanov}
\affiliation{ELI Beamlines Centre, Institute of Physics, Czech Academy of 
Sciences, Za Radnici 835, 25241 Dolni Brezany, Czech Republic}
\affiliation{National Institutes for Quantum and Radiological Science and 
Technology (QST), Kansai Photon Science Institute, 8-1-7 Umemidai, Kizugawa, 
Kyoto 619–0215, Japan}

\begin{abstract}
We show that an all-optical configuration of the laser-electron collision in 
the 
$\lambda^{3}$ configuration based on 10~PW-class lasers presents a viable 
platform for 
reaching the range of parameters where a perturbative QED in strong external 
electromagnetic field breaks. This case is contingently referred to as a case 
of 
the nonperturbative QED; and this range of parameters is the intriguing goal 
from 
an experimental point of view because of a possible manifestation of a new 
physics of the interaction of a highly radiating particle with a strong 
electromagnetic field.
We show that the strong field region can be reached by the electrons having 
the initial energy higher than 
50~GeV.
Our theoretical considerations are in agreement with three-dimensional 
particle-in-cell simulations.
While increasing of the electron energy raises the number of electrons 
experiencing the strong field region, the observable signature of photon 
emission radiative correction in the strong field is expected to fade out when 
the electron energy surpasses the 
optimal value.
This threshold of electron energy is identified and the parameters for achieving the nonperturbative limit of QED are provided.
\end{abstract}

\maketitle

The collision of laser-accelerated electrons with an intense laser pulse 
presents an all-optical configuration enabling the study of quantum 
electrodynamics (QED) effects such as photon emission and electron-positron 
pair generation 
\cite{Koga2005,DiPiazza2012,Vranic2016Quantum,Blackburn2020,Gonoskov2021,Cole2018,Poder2018}.
Further increasing of laser intensities and energies of laser-accelerated 
electrons  will bring us to the regime where the current perturbative theoretical 
framework of QED becomes inapplicable \cite{Ritus1972,Narozhny1980,Fedotov2017}.

The latter point demands clarification from a terminological point of view. 
A number of results obtained previously in the field of dynamics of photons and 
strongly radiated electrons in a strong electromagnetic (EM) field are based on 
the 
Furry picture~\cite{Furry51}. 
The total quantum EM field $\hat{\cal A}$ is represented as a sum of two 
fields: 
$\hat{\cal A}=A\hat{\cal I}+\hat{\cal A}_1$. 
Here $ \hat{\cal I} $ is the identity operator and $A$ is a four-potential 
of an external strong EM field. It is the classical 
field obeying the classical Maxwell equations. 
It is exactly (nonpertubatively) taken into account for `unperturbed' dynamics 
of real and virtual electrons and positrons. 
Relatively weak excitations over the unperturbed classic field are described 
by the quantum field $\hat{\cal A}_1$. 
Its presumably weak interaction with electrons and positrons in the strong 
field $A$ is described by the term $-\hat{j}\cdot\hat{\cal A}_1$ in a 
Lagrangian, where $\hat{j}$ is the quantum electric four-current. 
The Furry picture implies using the usual QED perturbation approach relative 
to this interaction, and, in particular, the perturbation approach (with the 
re-normalization) for any virtual photons, electrons and positrons  whose 
energies are much larger than the energies of real in-coming and out-going 
photons, 
electrons and positrons taking part in the process. 
The success of the latter procedure is provided by the smallness of the fine 
structure constant, $\alpha=e^2/(\hbar c)$. 
Here $e$ stands for the elementary electric charge, $c$ speed of light, and 
$\hbar$ is the reduced Planck constant.
This approach with its limitations are recently reviewed in detail in 
Ref.~\cite{Fe22}.
It appears that this approach cannot be applied in the case when energies of 
the real particles taking part in a process become very large, regardless of the
smallness of $\alpha$.
The formulation of this limitation will be considered in the next 
paragraph. 
However, we may say now that it originates from the fact that cascading 
creation of virtual electrons, positrons and photons with energies comparable 
with the energies of the real particles, when the energies become sufficiently 
large, destroys convergence of the perturbation theory considered above.
The range of the parameters, where such convergence breaks, will be 
called as a domain of `nonperturbative  QED'.

The interaction of an electron with the EM field is 
characterized 
by the Lorentz 
invariant parameter $ \chi_e=\sqrt{-\left( F^{\mu\nu}p_{\nu}\right)^{2} 
}/\left( m_{e}cE_{\mathrm{S}}\right)  $ where $ 
F_{\mu\nu}=\partial_{\mu}A_{\nu}-\partial_{\nu}A_{\mu}$ is the EM field tensor, 
$ A_{\nu} $ is the four-potential, $ p_{\nu} $ is the four-momentum of the 
electron,  $E_{\mathrm{S}} =m_{e}^{2}c^{3}/\left( e\hbar\right) \approx1.33\times10^{16}~\mathrm{V/cm} $ is the QED critical field and
$ m_{e} $ is  the electron mass \cite{Sauter1931,Schwinger1951,Ritus1985}.
For example, $ \chi_{e}\simeq0.1 $ has been reached in experiments presenting  
evidence of radiation reaction in the
collision of a high-intensity laser beam with laser-wakefield
accelerated electrons in the counter-propagating setup \cite{Cole2018,Poder2018}.
While for $ \chi_{e}\ll1 $ the quantum effects on the electron-laser 
interaction are negligible, they become important as $ \chi_{e}\gtrsim 1 $ 
\cite{Bulanov2011,DiPiazza2012,Blackburn2020,Gonoskov2021}.
%
In this case, the probabilistic photon emission as well as the recoil and 
straggling effects should be taken into account \cite{Harvey2017}.
Meanwhile, according to Ritus-Narozhny conjecture, the perturbative QED 
theoretical 
framework in a strong EM field fails due to the large radiative corrections when $ \alpha 
\chi_{e}^{2/3}\approx1 $ which corresponds to $ \chi_{e}\approx1600 $ 
\cite{Ritus1972,Narozhny1980,Fedotov2017}.
Therefore, several schemes have been proposed to experimentally probe this regime of QED.
However, reaching the highest values of $ \chi_{e} $ is possible only if the 
radiative losses are mitigated.
This can be done by reducing the interaction time between the electron and the EM field.
In the limit $ \chi_{e}\gg 1 $, the characteristic radiation time between the 
emission of two photons is $ \sim \gamma_{e}\hbar/( \alpha 
\chi_{e}^{2/3}m_{e}c^{2})  $ where $ \gamma_{e} $ is the relativistic Lorentz 
factor of the electron \cite{Ritus1985}.
If one considers the optical lasers, the spatial extent of the interaction is 
usually characterized by the laser wavelength $ \lambda \approx 1~\mathrm{\mu 
m}$, thus reaching $ \alpha \chi_{e}^{2/3}\approx 1$ seems to be hard 
using 100 GeV-class electron bunch as the interaction time is ten times longer 
than the characteristic radiation time \cite{Yakimenko2019,Baumann2019}.
Therefore other schemes reducing the interaction time have been put forward.
The concept suggested in Ref.~\cite{Baumann2019} is based on the generation of 
isolated ultra-intense 150~as pulse by the reflection of the optical laser 
pulse from a plasma mirror 
\cite{anderBrgge2010,Bulanov2003,Naumova2004,Vincenti2019} and its collision 
with a counterpropagating 125~GeV electron beam.
The radiation losses of counter-propagating electrons can be also mitigated by 
the appropriate choice of the collision angle \cite{Blackburn2019} or by 
employing a solid target \cite{Baumann2019Laser}.
Another approach considers the collision of two 125~GeV tightly compressed and focused electron bunches having longitudinal dimension $ < 100~\mathrm{nm}$ \cite{Yakimenko2019}.
It has been also shown that the limit of nonperturbative QED can be approached in the collision of $ \sim $TeV electrons with crystal atoms aligned along the symmetry direction \cite{DiPiazza2020}.

Here we consider the interaction of the 100~GeV-class electron beam with tightly-focused azimuthally 
polarized optical laser pulse in the $ \lambda^{3} $ configuration.
Contrary to linear or circular polarization, the azimuthal one provides an 
order of magnitude larger area of strong-field region (doughnut-shaped) and 
therefore it affects more particles. 
We show both analytically and numerically by means of full-scale 3D Particle-In-Cell (PIC) simulations that even though the laser pulse duration is an order of magnitude greater than the characteristic radiation time, using an optical 10~PW-class laser allows approaching the nonperturbative limit of QED.
The effect of photon emission correction due to the electron mass modification 
is discussed.
As this effect would manifest itself only if electrons with large $ \chi_{e} $ radiate, we expect that the signature of such a radiative correction on the final electron distribution starts to fade out 
once the electron initial energy surpasses the optimal value.

\begin{figure}[ht]
	\centering
	\includegraphics[width=1.0\linewidth]{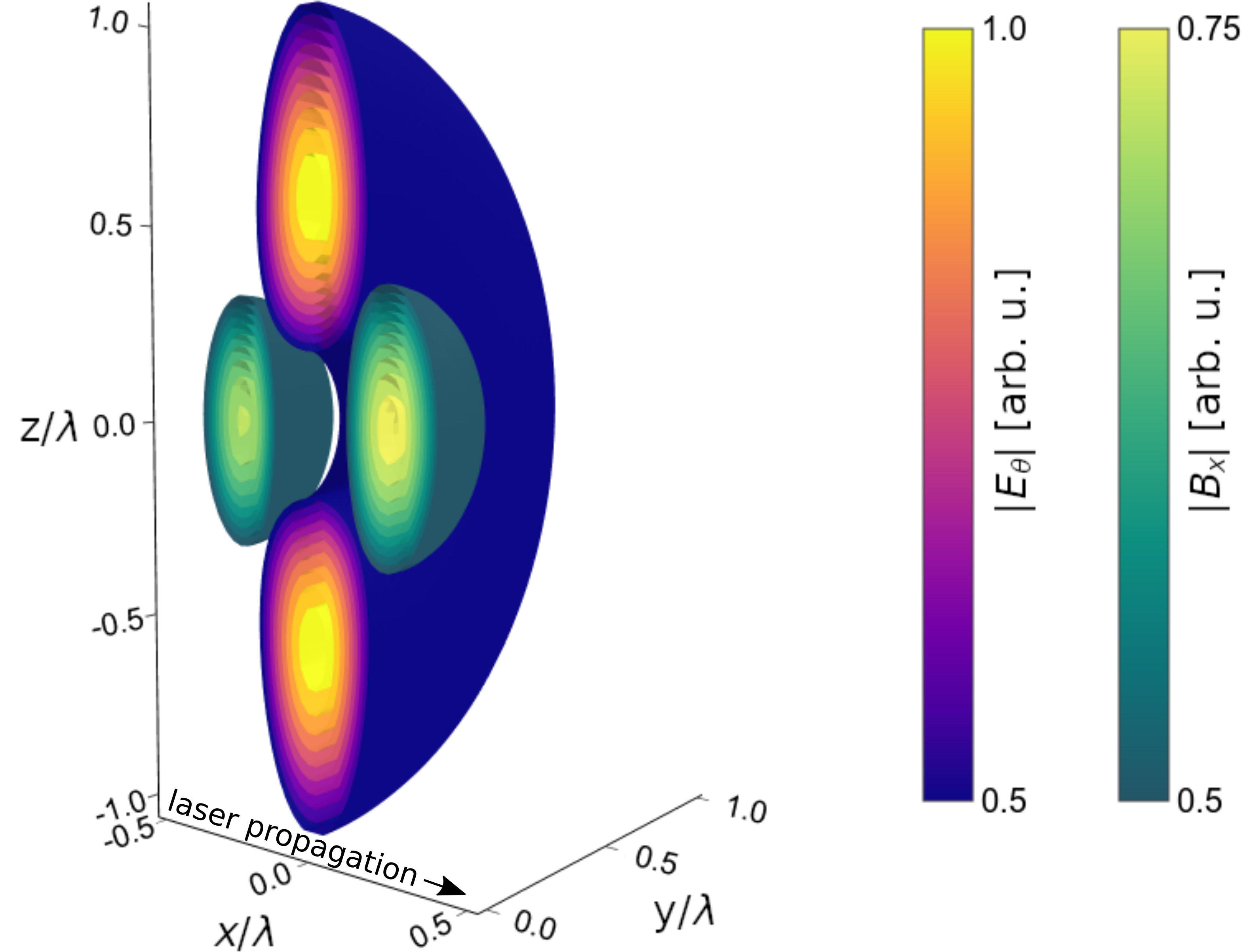}
	\caption{The field isosurfaces $\lvert E_{\theta}(x,y,z,t) \rvert$ and $ 
	\lvert B_{x}(x,y,z,t) \rvert $ of the laser pulse at $ t=0~\mathrm{s} $ for 
	the initial phase $ \varphi_{0}=0 $. The focal plane is at $ 
	x=0~\mathrm{\mu m} $. The laser pulse propagates in the positive $ x 
	$-direction.}
	\label{fig:01}
\end{figure}

The key factor for achieving high $ \chi_{e} $ with near-future 10~PW-class 
laser systems is to provide a strong EM pulse within a tiny spatio-temporal 
region.
Therefore we have considered the tightly focused azimuthally polarized laser 
beam expressed in high-order expansions of a parameter $ \epsilon=\lambda/\left( \pi w_{0}\right) $, where $ 
w_0 
$ is the waist radius of the Gaussian beam \cite{Salamin2006}:
\begin{widetext}
\begin{equation}\label{Br}
B_{r}\left(x,y,z,t \right) =E_{0}e^{-r^{2}/w^{2}} \left[ \epsilon\rho 
C_{2}+\epsilon^{3}\left(-\dfrac{\rho 
C_{3}}{2}+\rho^{3}C_{4}-\dfrac{\rho^{5}C_{5}}{4} 
\right)+\epsilon^{5}\left(-\dfrac{3\rho 
C_{4}}{8}-\dfrac{3\rho^{3}C_{5}}{8}+\dfrac{17\rho^{5}C_{6}}{16}-\dfrac{3\rho^{7}C_{7}}{8}+\dfrac{\rho^{9}C_{8}}{32}
 \right)  \right],
\end{equation}

\begin{equation}\label{Bx}
B_{x}\left(x,y,z,t \right)=E_{0}e^{-r^{2}/w^{2}}\left[\epsilon^{2}\left( 
S_{2}-\rho^{2}S_{3}\right) + \epsilon^{4}\left( 
\dfrac{S_{3}}{2}+\dfrac{\rho^{2}S_{4}}{2}-\dfrac{5\rho^{4}S_{5}}{4}+\dfrac{\rho^{6}S_{6}}{4}\right)
   \right],
\end{equation}

\begin{equation}\label{Etheta}
E_{\theta}\left(x,y,z,t \right)=E_{0}e^{-r^{2}/w^{2}}\left[\epsilon\rho 
C_{2}+\epsilon^{3}\left(\dfrac{\rho 
C_{3}}{2}+\dfrac{\rho^{3}C_{4}}{2}-\dfrac{\rho^{5}C_{5}}{4} \right) 
+\epsilon^{5}\left( \dfrac{3\rho C_{4}}{8} + \dfrac{3\rho^{3}C_{5}}{8} + 
\dfrac{3\rho^{5}C_{6}}{16} - 
\dfrac{\rho^{7}C_{7}}{4}+\dfrac{\rho^{9}C_{8}}{32}\right) 
   \right],
\end{equation}
\end{widetext}
where for $ n=2,3, ... $
\begin{equation}
C_{n}=\left( \dfrac{w_{0}}{w}\right) ^{n}\cos\left( \varphi+n\varphi_{G} 
\right),
\end{equation}
and
\begin{equation}
S_{n}=\left( \dfrac{w_{0}}{w}\right) ^{n}\sin\left( \varphi+n\varphi_{G} 
\right).
\end{equation}
The radial distance from the beam axis is $ r=\sqrt{y^{2}+z^{2}} $, $ \rho=r/w_{0} $, $ 
w=w_{0}\sqrt{1+\left( x/x_{R}\right)^{2} } $, $ x_{R}=kw_{0}^{2}/2 $ is 
the Rayleigh range, $ k=2\pi/\lambda $ is the wavenumber, $ 
\varphi=\varphi_{0}+\omega_{0} t - kx - kr^{2}/2R $, 
$\omega_{0}$ is the laser frequency, $ 
R=x+x_{R}^{2}/x $, $ \varphi_{0} $ is the initial phase and $ 
\varphi_{G}=\arctan\left( x/x_{R}\right)  $ is the Gouy phase.
We assume that the laser pulse is propagating along the positive $ x $-axis.
For $ \varphi_{0}=0 $, the  radial magnetic $ B_{r} $ and azimuthal electric $ E_{\theta} $ 
fields reach their maxima at time $ t=0~\mathrm{s} $ in the focal plane $ 
x=0~\mathrm{\mu m} $, while the longitudinal magnetic field $ B_{x} $ is zero 
here, see Fig.~\ref{fig:01}.
The peak power (with respect to the peak intensity) of a tightly focused beam 
is 
given by \cite{Salamin2006}
\begin{equation}
\mathcal{P}=\dfrac{\pi w_{0}^{2}}{2}\dfrac{B_{x}^{2}}{\mu_{0}}\left( 
\dfrac{\epsilon}{2}\right)^{2} \left[ 1+3\left( 
\dfrac{\epsilon}{2}\right)^{2}+9\left( \dfrac{\epsilon}{2}\right)^{4} \right].
\end{equation}

The laser field is characterized by the parameter 
$a_{0}=eE_{0}/(m_e\omega_{0} c) $, where $ E_{0} $ is the amplitude of 
the electric field.
Assuming $ \gamma_{e}\gg 1 $, the maximum $\chi_e$ that can 
be achieved in a head-on collision is $2\gamma_{e}E_{0}/E_{\mathrm{S}} $.
However, it is not guaranteed, that the maximum value of $ \chi_{e} $ parameter 
will be achieved by the radiating particle as this depends on laser parameters 
and electron initial energy.
Therefore, below we estimate the ratio of the initial electron number, that achieves the 
laser pulse center without radiating a photon.

In the case of the azimuthally polarized laser beam, the $ \chi_e $ parameter 
is 
maximized when the electron interacts with the maximum of the 
counterpropagating 
EM field created by the radial magnetic $ B_{r} $ and 
azimuthal electric $ 
E_{\theta} $ 
fields.
These transverse fields generate a 
doughnut-shaped intensity structure.
Since the EM field structure of the azimuthally polarized laser pulse is 
cylindrically symmetric, we analyze the interaction with an electron in 1D 
geometry taking advantage of the approach described in Ref.~\cite{Jirka2021}.
However, in the tightly focused laser beam, the azimuthal component of the 
field is 
always stronger than the radial one and they reach their maxima at different 
radial distances.
To take this effect into account, we consider the average strength of the 
transverse 
electromagnetic field 
\begin{equation}\label{a0perp}
a_{0\perp}=\dfrac{eG_{\perp}}{m_e\omega_{0} c},
\end{equation}
where  
$G_{\perp}= \mathrm{max}\left(\lvert B_{r}+E_{\theta}\rvert/2 \right)  $.
The maximum expected value of the corresponding $ \chi_{e} $ parameter is, therefore, 
\begin{equation}
\chi_{e\perp}=2\gamma_{e} \dfrac{a_{0\perp}}{a_{\mathrm{S}}},
\end{equation}
where $ a_{\mathrm{S}} $ corresponds to the QED critical field.

We assume that the electron interacts with a laser pulse having a Gaussian 
temporal envelope of full-width-at-half-maximum 
(FWHM) duration $ \tau $ in laser intensity.
In the limit $ \chi_{e} \gg 1 $, the average energy of the electron 
in the center of the laser pulse of a FWHM duration $\tau=T/2$, where $ T $ is the laser period, can be approximated 
as
\begin{equation}\label{E_e_c}
\mathcal{E}_{e}^{\mathrm{c}}\approx 
\left( 1-16/63\right)^{W_{\gamma}\tau/\sqrt{2\ln 2}}\mathcal{E}_{e},
\end{equation}
where 
$ \mathcal{E}_{e} $ is the initial electron energy, and
\begin{equation}\label{W}
W_{\gamma}\approx \dfrac{3^{2/3}28\Gamma\left( \frac{2}{3}\right)\alpha 
m_{e}^2c^{4}\chi_{e\perp}^{2/3}}{54\hbar\mathcal{E}_{e}}
\end{equation} is the rate of 
single-photon emission per unit time for the Compton process and $ \Gamma\left( 
x\right)  $ is the Gamma function \cite{Ritus1985,Jirka2021}.
The factor 16/63 presents the average energy of the emitted photon 
\cite{Ritus1985} and the 
exponent 
defines the number of emitted photons.
Due to the relationship between the Gaussian and flat-top temporal envelope, 
the 
effective amplitude is $ a_{0\perp}/\sqrt{2} $ resulting in the effective $ 
\chi_{e} $ 
parameter
$\sqrt{2}\gamma_{e}a_{0\perp}/a_{\mathrm{S}} $ and the full pulse 
length $ \tau $ gets a factor of $ 2/\sqrt{2\ln 2} $.
Therefore the time required for an electron to reach the center of the 
laser pulse is $ \tau/\sqrt{2\ln 2}. $
%
%

%
If such an electron radiates, it does not reach the maximum $ \chi_{e} $ 
parameter in the center of 
the laser pulse.
%
Therefore $ \mathcal{E}_{e}^{\mathrm{c}}/\mathcal{E}_{e} $ 
approximates the 
ratio of the number of electrons that achieve the laser pulse center without 
emitting a 
photon (and thus reach the maximum $ \chi_{e} $) to the total number of 
electrons.
To maximize this ratio for a fixed interaction duration $ \tau $, the rate $ 
W_{\gamma} \propto a_{0\perp}^{2/3}/\gamma_{e}^{1/3} $ has to be reduced while 
keeping $ 
\alpha \left(2\gamma_{e}a_{0\perp}/a_{\mathrm{S}} \right) ^{2/3}\approx 1 $ in 
order to achieve the nonperturbative limit of QED.

\begin{figure}
	\centering
	\includegraphics[width=1.0\linewidth]{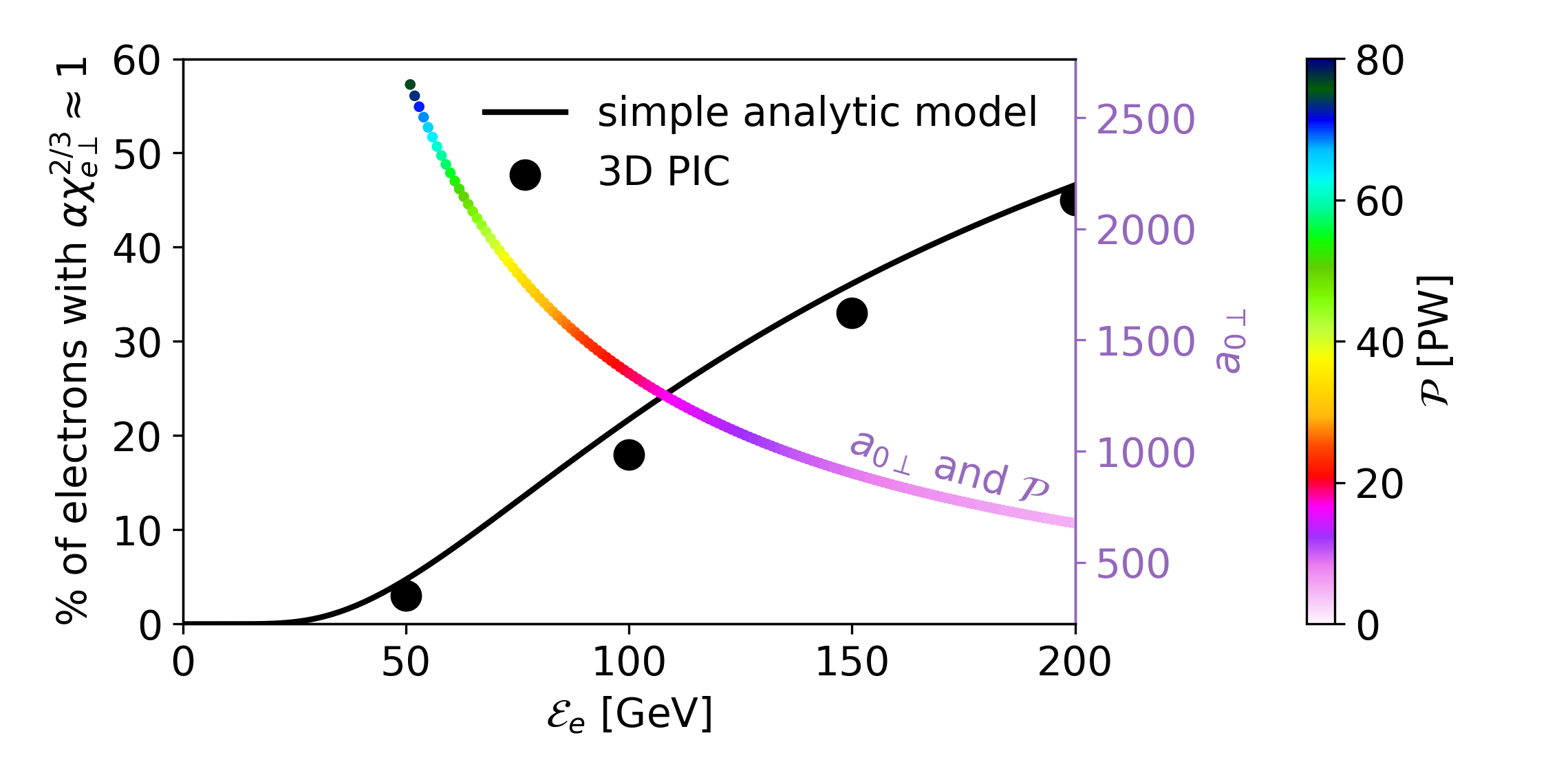}
	\caption{The ratio of electron number reaching $ 
	\alpha\chi_{e\perp}^{2/3}\approx1 $ 
	as a function of their initial energy $ \mathcal{E}_{e} $ for laser 
	parameters mentioned in the text. The black solid line represents the 
	expected value obtained from Eq.~\eqref{E_e_c}. The 
	bullets 
	represent the results from 3D PIC simulations. The second line shows the 
	corresponding $ a_{0\perp} $ (right axis) and its color the required power 
	$ \mathcal{P} $.}
	\label{fig:02}
\end{figure}

In the following, we consider the $ 
\lambda^{3} $ configuration of the laser-electron collision \cite{Mourou2002}.
The spatio-temporal parameters of the azimuthally polarized laser pulse are as follows:
$\lambda = 0.8~\mathrm{\mu m} $, duration $ \tau=T/2 $ and $ 
w_{0}=0.424\lambda $.
The initial electron energy $ \mathcal{E}_{e} $ is in the range 
50--200~GeV.
The results for the above-mentioned parameters are shown in 
Fig.~\ref{fig:02}.
The black solid line represents the simple analytic model estimating the ratio 
of the 
number of electrons that do not emit before reaching the laser pulse center to 
their total number.
The corresponding $ a_{0\perp} $ is depicted by the second line colored according to the laser peak power $ \mathcal{P} $.
For e.g. 10~PW laser, the nonperturbative limit of QED is reached in the head-on collision with 140~GeV electrons.
In this case, 1/3 of electrons interacting with the region of a strong 
transverse electromagnetic field $ a_{0\perp}\approx960 $ reach $ \alpha 
\chi_{e\perp}^{2/3}\approx1 $.
The energy of laser-wakefield accelerated electrons scales as $ 
\mathcal{E}_{e}\left[\mathrm{GeV} \right] \propto 
10\times\mathcal{P}\left[\mathrm{PW} \right]  $, see Ref.~\cite{Bulanov2016} 
and references cited therein.
The
combination of 
10~PW-class lasers thus provides the all-optical scheme for probing the 
nonperturbative 
limit of QED.

We have benchmarked our analytical results against 3D PIC simulations of 
laser-electron collision in the code {\sf SMILEI}
in which photon emission 
and Breit-Wheeler
electron-positron pair creation are implemented via the Monte-Carlo method  \cite{Derouillat2018}.
The simulation box dimensions are $ 4\lambda\times20\lambda\times20\lambda $ 
resolved with $ 512\times2560\times2560 $ cells.
The EM fields given in Eqs.~\eqref{Br}--\eqref{Etheta} are 
numerically evaluated at the simulation box boundary for the whole duration of 
the laser pulse and they propagate towards 
the 
center of the simulation box by using the finite-difference time-domain Maxwell 
solver \cite{Derouillat2018}.
The tightly focused laser pulse propagates in the positive $ x $-direction and 
collides head-on with $ 10^{9} $ electrons distributed in a $ \lambda/5 $ 
thick disc of a radius $ \lambda $ initially located at a such distance $ x $ 
that it reaches the focal plane $ x=0~\mathrm{\mu m}$ at $ t=0~\mathrm{s}$.
The electron initial energy $ \mathcal{E}_{e} $ is in the range 50--200~GeV.
The parameters of the laser pulse are the same as in the previous text.
Since $ a_{0}\gg 1 $, the condition for local constant field approximation is satisfied \cite{Ritus1985,Kirk2009}.
From the simulation data one can obtain the $ \chi_{e} $ parameter of each particle and thus identify the ones that approach the nonperturbative limit of QED.
Their ratio is obtained by comparing the number of particles that $ \chi_{e} $ 
reach at least $ \alpha \chi_{e}^{2/3}=0.95 $ with the number of all particles 
originating in the same region.
As can be seen in Fig.~\ref{fig:02}, the simple analytic estimate (black line) 
well 
approximates the results obtained in 3D simulation (bullets) across the whole 
studied energy range.

We see in Fig.~\ref{fig:02} that the higher the initial electron energy, 
the more particles 
approach the nonperturbative limit of QED.
As in Refs.~\cite{Yakimenko2019,Baumann2019,Baumann2019Laser}, we 
have used the model of the photon emission 
rate which has been developed within the perturbative theory in order to 
perform the predictive simulations for reaching the nonperturbative limit of 
QED.
Therefore, applying this approach for $ \alpha 
\chi_{e\perp}^{2/3}< 1 $ would increase the validity of the predictions. 
At the nonperturbative limit $ \alpha 
\chi_{e\perp}^{2/3}= 1 $, the exact theory taking into account all the 
radiative 
corrections 
to electron and photon motion is needed \cite{Ritus1970,Ritus1972}.
For example, it is known that the existence of photon emission in a 
strong EM field is 
connected with the 
radiative effect leading to the electron mass increase \cite{Ritus1970}.
This electron mass shift may consequently affect the photon emission 
rate. 
However, the 
exact calculation of the corresponding photon emission rate 
correction is beyond 
the scope of this paper.

Using the approach described in Ref.~\cite{Jirka2021} one can calculate the 
average final electron energy while neglecting the radiative corrections.
This value might serve as a reference for further theoretical and experimental 
discussions considering the radiative corrections.
If the electron mass modification would not be neglected in the 
calculation of photon emission rate, we expect that the modified rate would 
result in a final electron energy distribution that differs from the one 
calculated by Eq.~\eqref{W}.
Therefore it is interesting to inspect the question what are the optimal 
parameters of laser-electron collision for observing the effect of a modified 
photon emission rate?
We note, that the effect of mass modification on photon emission rate has 
been previously considered in Ref.~\cite{Yakimenko2019} within the perturbative 
QED 
framework.
In this cited paper, the magnitude of a correction to the photon emission rate 
resulting from  the 
change of the electron mass was estimated by considering the corresponding 
change in $ \chi_{e} $  parameter.

In general, the effect of a correction in a photon emission rate 
on the final 
electron 
energy can only manifest itself if electrons 
having large $ \chi_{e} $ parameter
radiate.
This poses contradictory requirements on the laser-electron interaction as the 
maximum $ \chi_{e} $ can be reached by the electron only 
if photon emission 
is 
mitigated.
In other words, there is a trade-off between using high electron energy to 
reach high $ \chi_{e} $ and maximizing the amount of emitted energy.
If the energy of the colliding electron is relatively low, it will radiate all 
its 
energy before reaching the high value of $ \chi_{e} $.
The higher values of $ \chi_{e} $ can be achieved with the increasing initial 
electron 
energy, however, the photon emission rate is proportional to $ 
1/\gamma_{e}^{1/3} $, 
thus less amount of its initial energy would be emitted.
And finally, if the energy of the colliding electron is high enough, it 
will reach the center of the laser pulse without emission and therefore achieve 
the maximum value 
of $ \chi_{e} $.
Therefore, we need to optimize the electron and laser parameters in order to 
maximize the 
energy emitted by the electrons with a high $ \chi_{e} $ parameter.

The amount of energy emitted by the electron during its propagation 
to the center of the laser pulse is on average given by 
$\mathcal{E}_{e}-\mathcal{E}_{e}^{\mathrm{c}}$.
We are interested in the emission of electrons with a high $ \chi_{e} $ parameter.
The parameter $ \chi_{e} $ in Gaussian temporal envelope gradually increases from 0 to 1600.
At a distance FWHM/2=$ \tau/2 $ from the center of the laser pulse, the value of $ \chi_{e}=1600/\sqrt{2}\approx1100 $ is reached.
In the following, we consider "low" ($ \chi_{e} \leq 1100 $) and "high" ($ 1100 < \chi_{e} \leq 1600 $) values of $ \chi_{e} $ parameter.
The amount of energy emitted by electrons with "high" $ \chi_{e} $ parameter can be obtained as the total emitted energy $ \left( \mathcal{E}_{e}-\mathcal{E}_{e}^{\mathrm{c}} \right)  $ minus energy emitted by electrons with "low" $ \chi_{e} $ parameters. 
Due to the geometrical correspondence between the Gaussian and the flat-top temporal envelope, the amount of energy emitted by electrons with "low" $ \chi_{e} $ parameter can be calculated by only considering the corresponding contraction of the flat-top laser pulse duration.
For this case, the duration of a flat-top laser envelope has to be
modified by a factor of 0.25.
Thus the energy delivered by the front part of the Gaussian laser pulse 
up to a moment of reaching $ \chi_{e}=1100 $ corresponds to a flat-top temporal 
envelope of duration $\approx \tau/4\sqrt{2\ln 2} $.
Therefore, the amount of energy emitted by electrons with "low" $ \chi_{e} $ is
\begin{equation}\label{E_e_c_low}
\mathcal{E}_{e}-\mathcal{E}_{e}^{\mathrm{low~}\chi_{e}}= \mathcal{E}_{e}-
\left( 1-16/63\right)^{W_{\gamma}\tau/4\sqrt{2\ln 2}}\mathcal{E}_{e}.
\end{equation}
Then
\begin{equation}\label{E_e_c_high}
\Delta \mathcal{E}_{e}^{\mathrm{high~}\chi_{e}}\approx \left( \mathcal{E}_{e}- \mathcal{E}_{e}^{\mathrm{c}}\right)   - \left( \mathcal{E}_{e}- \mathcal{E}_{e}^{\mathrm{low~}\chi_{e}} \right)
\end{equation}
characterizes the amount of energy emitted by electrons with "high" $ 
\chi_{e} $.
%
%
%
%
%
To maximize this quantity, we define  
\begin{equation}\label{f}
f\left( \gamma_{e}\right) 
=\dfrac{\Delta \mathcal{E}_{e}^{\mathrm{high~}\chi_{e}}\left( \gamma_{e}\right) }{\mathcal{E}_{e}\left( \gamma_{e}\right) }=\dfrac{\mathcal{E}_{e}^{\mathrm{low~}\chi_{e}}\left( \gamma_{e}\right) 
-\mathcal{E}_{e}^{\mathrm{c}}\left( \gamma_{e}\right) 
}{\mathcal{E}_{e}\left( \gamma_{e}\right) }
\end{equation}
as it represents the ratio of the initial electron energy that is radiated by 
electrons having a "high" $ \chi_{e} $ parameter.
%
%
%
%
%

We propose to use the average final electron energy as the observable 
quantity 
which could be altered as a consequence of photon emission rate correction.
To make this difference noticeable, we need to maximize the amount of energy 
emitted by 
electrons with a 
high 
value of $ \chi_{e} $ parameter.
The reason is that the radiative corrections are expected to become 
non-negligible for 
electrons with high $ \chi_{e} $.
The radiative correction of photon emission rate can only be observable in the 
final electron energy distribution if many electrons of such a high $ \chi_{e} 
$ radiate.
The optimal initial electron energy for observing the signature of 
photon 
emission correction can be found by solving $ \left( \mathrm{d} f /\mathrm{d} 
\gamma_{e}\right)\vert_{\alpha \chi_{e\perp}^{2/3}=1} =0$.
For our parameters, $ f\left(\gamma_{e} \right)  $ 
reaches its maximum at $\gamma_{e}$ that corresponds to the initial energy of 
approximately
$
80~\mathrm{GeV} $.
This results indicates that the amount of energy radiated by electrons with 
high $ \chi_{e\perp} $ will be maximized if the interacting electrons will have 
the 
above-mentioned 
initial energy (while keeping $ 
\alpha \chi_{e\perp}^{2/3}=1 $ in 
this case).
%
%
Therefore, we expect that the effect of photon emission correction on 
the final 
electron energy will be most obvious for $ 
\mathcal{E}_{e}\approx80~\mathrm{GeV} $.
However, the exact calculation of the magnitude of the photon emission rate 
correction goes beyond the scope of this paper and needs further development of 
theoretical framework applicable in the range where the perturbative QED 
approach fails.
We provide the parameters for reaching the strong field region in the 
collision of an electron beam with a counter-propagating azimuthally polarized 
laser pulse.
We estimate how many electrons reach the nonperturbative limit of QED 
characterized by $ \alpha\chi_{e}^{2/3}\approx1 $.
When the interaction in the $ \lambda^{3} $ configuration is considered, the 
initial 
electron 
energy required for achieving the nonperturbative QED limit needs to be higher 
than 
50~GeV.
It is shown that e.g. for 10~PW laser, the nonperturbative limit of QED is 
experienced by one-third of the interacting electrons having energy 140~GeV, 
thus the
upcoming generation of 10~PW-class lasers provides a viable all-optical scheme 
for probing the nonperturbative QED.
The theoretical considerations are benchmarked against three-dimensional particle-in-cell simulations.
The effect of photon emission correction due to the electron mass modification is discussed.
We expect that the signature of such a radiative correction on final electron distribution starts to fade out 
once the electron initial energy surpasses the optimal value.

This work is supported by the project High Field Initiative (HIFI) 
CZ.02.1.01/0.0/0.0/15\_003/0000449 from European Regional Development Fund 
(ERDF).

\bibliography{ap}{}
\bibliographystyle{apsrev4-2}
\end{document}